\documentclass[final,1p,times,11pt,a4paper]{article}
\pdfoutput=1
\usepackage{hyperref}  
\hypersetup{pdfpagemode=None,pdfstartview=FitH}
\usepackage{amsmath}
\usepackage{graphicx}
\usepackage[numbers,super]{natbib}
\usepackage{color}
\usepackage{txfonts}
\usepackage{upgreek}
\usepackage{bm}
\usepackage{authblk}
\usepackage{textcomp}
\usepackage[font=normal,labelfont=bf]{caption}
\usepackage[a4paper, total={6.5in, 9.5in}]{geometry}
\usepackage{lineno, blindtext}
\usepackage[normalem]{ulem}

\newcommand{\invisible}[1]{}
\graphicspath {
    {./}
    {./Figures/}
}

\newcommand*\samethanks[1][\value{footnote}]{\footnotemark[#1]}

\begin{document}

\title{Swimming \emph{Euglena} respond to confinement with \\ a behavioral change enabling effective crawling}

\author[1]{Giovanni Noselli}
\author[2]{Alfred Beran}
\author[3,4]{Marino Arroyo\thanks{Corresponding authors\\ marino.arroyo@upc.edu\\ desimone@sissa.it}}
\author[1,5]{Antonio DeSimone\samethanks[1]}

\affil[1]{SISSA--International School for Advanced Studies, 34136 Trieste, Italy}
\affil[2]{OGS--Istituto Nazionale di Oceanografia e di Geofisica Sperimentale, 34151 Trieste, Italy}
\affil[3]{Universitat Polit\`ecnica de Catalunya-BarcelonaTech, 08034 Barcelona, Spain}
\affil[4]{Institute for Bioengineering of Catalonia (IBEC), The Barcelona Institute of Science and Technology, 08028 Barcelona, Spain}
\affil[5]{The BioRobotics Institute, Scuola Superiore Sant'Anna, 56127 Pisa, Italy}

\date{}
\setcounter{Maxaffil}{0}
\renewcommand\Affilfont{\itshape\small}

\maketitle


\begin{abstract}
\normalsize
Some euglenids, a family of aquatic unicellular organisms, can develop highly concerted, large amplitude peristaltic body deformations. This remarkable behavior has been known for centuries. Yet, its function remains controversial, and is even viewed as a functionless ancestral vestige. Here, by examining  swimming \emph{Euglena gracilis} in environments of controlled crowding and geometry, we show that this behavior is triggered by confinement. Under these conditions, it allows cells to switch from unviable flagellar swimming to a new and highly robust mode of fast crawling, which can deal with extreme geometric confinement and turn both frictional and hydraulic resistance into propulsive forces. To understand how a single cell can control such an adaptable and robust mode of locomotion, we developed a computational model of the motile apparatus of \emph{Euglena} cells consisting of an active striated cell envelope. Our modeling  shows that gait adaptability does not require specific mechanosensitive feedback but instead can be explained by the mechanical self-regulation of an elastic and extended motor system. Our study thus identifies a locomotory function and the operating principles of the adaptable peristaltic body deformation of \emph{Euglena} cells.\end{abstract}

\clearpage


\noindent Euglenids are a diversified family of unicellular flagellated protists abundant in a variety of aquatic ecosystems. Many species of euglenids are capable of performing large amplitude, elegantly coordinated cell body deformations. This behavior is referred to as the euglenoid movement or \emph{metaboly}.\cite{buetow1968biology,Euglenids-tree-of-life,handbook_protists} Depending on the species and within a species, metaboly can range from rounding and gentle bending or twisting  to periodic highly concerted peristaltic waves traveling along the cell body. Observations of metaboly date back to the first microscopists,\cite{Dobell} and have fascinated researchers across disciplines ever since. It is known that an active striated cell envelope, called pellicle, controls cell shape.\cite{buetow1968biology,suzaki1,suzaki3,Arroyo2012} However, the precise mechanism by which this cortical complex performs metaboly, including the molecular motors involved, is not known. More strikingly, the function of this behavior remains elusive. 

Metaboly has been interpreted as a mode of locomotion in a fluid environment.\cite{motility} In fact, the peristaltic version of metaboly has inspired prototypical models for low Reynolds number swimming,\cite{pushmepullyou} and theoretical studies have shown that, thanks to its non-reciprocal nature, it is a competent swimming strategy.\cite{Arroyo2012} Yet, since euglenids exhibiting metaboly are also capable of flagellar locomotion,\cite{Rossi2017} which allows them to move in a fluid about 50 times faster, swimming is not a compelling function justifying metaboly, a behavior requiring an extended and intricate cellular machinery.\cite{buetow1968biology,Euglenids-tree-of-life,handbook_protists} In euglenids feeding on large eukaryotic cells, it is accepted that metaboly enables the  large cell deformations required for phagocytosis. Since eukaryovorous ancestors have evolved into osmotrophic and autotrophic euglenids, which do not engulf large particles but still exhibit metaboly, it has been argued that metaboly in these species may be an ancestral vestige without a specific function.\cite{leander_2012,pellicle_evolution,anna} It has also been speculated that metaboly may be useful to break the protective cyst that some species can secrete and exit from it, to move in confined environments when some species penetrate dead animals or eggs to feed, or when other species lacking emergent flagella crawl in granular media.\cite{buetow1968biology,new_leander}  These hypotheses, however, have not been systematically examined.

\section*{Confinement triggers metaboly}

To investigate the role of confinement in the behavior of euglenids and their motility, we examined  cultures of \emph{Euglena gracilis}, a prototypical photosynthetic species exhibiting metaboly, at various degrees of crowding (Fig.~1a and Supplementary Movie~S1). In dilute cultures, cells exhibited fast swimming  in the posterior-to-anterior direction powered by an anterior flagellum, cruising at  $\sim$\,68\,$\upmu\text{m/s}$. During flagellar swimming, cells maintained a fixed cigar shape $\sim$\,50\,$\upmu\text{m}$ long and with a maximum diameter of $\sim$\,9\,$\upmu\text{m}$. Cells performed sharp turns as previously described.\cite{Jennings1904} In crowded cultures, presumably triggered by mechanosensation at the cell envelope and/or at the flagella,\cite{buetow1968biology} cells exhibited a variety of behaviors in addition to fast swimming, including rounding, bulging, and large amplitude periodic cell deformations. Rounding was often associated with cell spinning, whereas metaboly did not seem to have a significant effect in locomotion.

We also triggered metaboly by confining cells between two glass plates separated along one side  by a spacer to produce a wedge-shaped fluid chamber. Cells in the narrower regions of the chamber (with a typical gap of $\sim$\,5\,$\upmu\text{m}$) systematically developed large amplitude periodic shape changes. Using brightfield reflected light microscopy (Supplementary Methods), we were able to record simultaneously and continuously in time the shape changes and the reconfigurations of the pellicle envelope in the plane of the glass plate (Fig.~1b and Supplementary Movies~S2 and S3). The pellicle is composed of interlocking narrow and long  proteinaceous strips, often arranged helically and spanning from the anterior to the posterior ends of the cell. These  strips lie beneath the plasma membrane and are subtended by systems of microtubules.\cite{Euglenids-tree-of-life} Previous observations in another species (\emph{Euglena fusca}) showed that cell deformations were accompanied by sliding between adjacent pellicular strips, which kept constant their length and width,\cite{buetow1968biology,suzaki1,suzaki2} similarly to how flagellar shape depends on microtubule sliding in the axoneme.\cite{Lineaar1968} Inter-strip active sliding results in  local deformation of the cell envelope, a simple shear along the direction of the strips (Fig.~1c), which when coordinated in space and time can explain the shape dynamics during metaboly.\cite{Arroyo2012,Arroyo2014} 

Our continuous observations were consistent with this sliding model; strip width remained nearly constant ($\sim$\,560\,nm), and in agreement with a simple shear local deformation of the pellicle, the total cell area remained nearly constant during shape transformations (Supplementary Methods). To further examine the shape-morphing mechanism, we developed a theoretical model linking strip curvature on a planar surface, as in Fig.~1(b), and differential sliding along this strip, Fig.~1(d) and Supplementary Note 1. It allowed us to quantify over time the sliding displacement along selected strips, Fig.~1(e), finding net sliding inter-strip displacements of about one micron during the shape excursion in Fig.~1(b). By comparing sliding displacements at different instants, we estimated sliding velocities of up to $\sim$\,1\,$\upmu\text{m/s}$, compatible with those of molecular motors along microtubules.\cite{Bio_numbers} Besides confirming the pellicle strip sliding model, these observations showed that cell volume was nearly constant during metaboly (Supplementary Methods). Interestingly, cells performing metaboly confined between two parallel plates exhibited directed motion in the  anterior-to-posterior direction at speeds ranging between 0.08 and 0.2 body lengths per cycle (Supplementary Movies~S2 and S3).

To mimic the more realistic situation of multidimensional confinement while having a high degree of experimental control, we drove swimming cells towards tapered glass capillaries with diameter ranging between 300\,$\upmu$m and 6\,$\upmu$m with slopes always smaller than 0.006 in the region of interest, Fig.~1(f) and Supplementary Fig.~2. When the diameter of the capillary was significantly larger than the cell diameter, cells exhibited fast swimming. As cells moved towards the narrow end of the capillary, they collided with the wall and started developing the behaviors described in the crowded cultures, including bulging and rounding, Supplementary Movie~S4. By rounding, some cells were able to switch their orientation in the capillary, Supplementary Movies~S4 and S5. Most cells eventually developed characteristic highly coordinated and periodic peristaltic body deformations consisting of a bulge moving in the posterior-to-anterior direction followed by a recovery phase during which the bulge at the posterior end reappears at the expense of that at the anterior end, Fig.~1(f)iv and Supplementary Movie~S4. Thus, similarly to culture crowding, confinement provided by the capillary triggered changes in cell behavior, which now seemed to acquire a function: cell rounding to turn and metaboly to crawl.

\section*{Metaboly is an effective crawling strategy}

Close examination of cells performing metaboly in the capillary showed that during their gait, the bulge moving in the posterior-to-anterior direction transiently contacted the capillary wall propelling the cell in the anterior-to-posterior direction, opposite to that of flagellar swimming, Fig.~2(a)i and Supplementary Movie~S6. The kinematics of the gait were highly non-reciprocal, and thus in principle compatible with self-propulsion in the non-inertial limit of our experiments.\cite{Purcell-1977-a} Since cells could switch direction in the capillary, we observed cells crawling by metaboly away from and towards its narrower end. As a result, we were able to examine the features of this mode of locomotion at varying degrees of confinement. To ease understanding of pellicle kinematics, we arranged images in the figures and movies so that cells crawl from left to right.  At higher confinement, cells adapted their gait by developing a broader bulge, which remained in contact with the wall during a larger fraction of the gait, Fig.~2(a)ii and Supplementary Movie~S6. However, the essential features of the crawling mechanism remained the same. 

Kymographs of cells crawling showed the high regularity of the gait and its adaptation to increasing confinement, Fig.~2(b). They also revealed that the net cell displacement was the result of a power phase, when the bulge travels towards the anterior end, and a recovery phase with backward cell motion.  Remarkably, cells were able to crawl up to very high  degrees of confinement, despite the little space available for shape transformations. Cells crawled fastest, at about 2.5\,$\upmu$m/s (0.4 body lengths per cycle), at intermediate degrees of confinement, where the bulge established sustained contact with the capillary walls and there was sufficient space for significant shape excursions, Fig.~2(c). The period of the gait, however, was largely independent of confinement, Fig.~2(d).   
Interestingly, the very small crawling velocities we recorded for weakly confined cells (Fig.~2(a) and largest diameter in Fig.~2(c)), which represent a transition between crawling and swimming, are in agreement with a theoretical study examining the complementary situation of swimmers undergoing large shape changes near walls under weaker capillary confinement.\cite{Wu2016}

We then wondered if flagellar locomotion of fixed-shape cells would be effective in this situation. Confined cells propelled by flagella would have to overcome a frictional force against the capillary walls and a hydraulic resistance. We theoretically estimated that such cells would move 10 to 20 times slower than those crawling by metaboly, Supplementary Note 3. These estimations were consistent with observations of stuck cells beating their flagella, Supplementary Movie~S8. Thus, by developing metaboly under confinement, cells switched from ineffective flagellar propulsion to a highly robust crawling mode of locomotion.

\section*{Mechanism for locomotion during metaboly}

To propel their body forward during migration in a low Reynolds number limit, \emph{Euglena} cells must exert self-equilibrated forces on their environment, here, the walls and the fluid within the capillary. To experimentally characterize the physical interaction between cells and their environment, we drove cells not exhibiting deformations into narrow sections of capillaries by applying a known pressure difference while recording cell velocity, contact area, and motion of suspended beads, Supplementary Movie~S7. These observations established that confined cells acted as hydraulic plugs and characterized the friction between cells and the wall as viscous and confinement-dependent, Supplementary Note 2.

We attempted to understand force transmission during crawling by metaboly in the light of a prevalent model of animal cells crawling in narrow spaces. According to this model, non-adherent confined cells generate propulsive forces through retrograde actin flows and unspecific friction with the confining wall. This propulsive force is balanced by a resistive hydrodynamic force required to displace the water column in the capillary,  \cite{Hawkins2009,Hawkins2011,Bergert2015,Prentice-Mott2013,Liu2015} unless water transport is sufficiently fast across the cell.\cite{new_hydraulic} This framework has explained why a large hydraulic resistance relative to the wall friction stalls cell motion and results in fast retrograde flow, whereas vanishing hydraulic resistance leads to fast cell motion and minimal sliding between the polarized actin cytoskeleton relative to the wall. Transposing this model to \emph{Euglena} cells crawling by metaboly, the backward motion of the pellicle bulge would be the analog of retrograde actin flow of animal cells, Fig.~3(a).

To test this analogy, we developed an idealized theoretical model of the power phase of the gait consistent with the shape-morphing principle of the pellicle of euglenids,\cite{Arroyo2012,Arroyo2014}  Fig.~3(b). In this model, a bulge establishes contact with a capillary of radius $r$ over a length $\ell_c$, and moves backwards at speed $c<0$ in the reference frame of the cell, which otherwise is a cylinder of radius $r_0$. The strips in the contact region form an angle $\theta^*$ with the cell axis satisfying $\cos \theta^* = r_0/r$. To weigh the relative importance of hydraulic and frictional resistance, we introduced the non-dimensional quantity $\xi = \mu_{\rm wall} \ell_c/(\alpha r)$, where $\mu_{\rm wall}$ is the wall-cell friction coefficient and $\alpha$ is a hydraulic resistance coefficient, Supplementary Note 4.

We first considered the limit of vanishing hydrodynamic resistance relative to wall friction, expressed as $\xi\rightarrow +\infty$. In this situation, the cell velocity $v = -(1-\cos\theta^*)\, c$ was obtained from the kinematics of the propagating bulge  by requiring no slippage between the pellicle and the wall, Fig.~3(c)i and Supplementary  Movie~S9. This expression is consistent with the very small cell velocities at extreme confinement, where $r\approx r_0$ or $\cos\theta^*\approx 1$, Fig.~2(c). To examine the effect of cell motion on the fluid inside the capillary in this no-slip scenario, we evaluated with our model the induced water flow rate $Q$ in either side of the bulge, which we assumed to act as a hydraulic plug. Strikingly, we found that as cells crawl without sliding in a given direction, they pump water in the opposite direction at a flow rate given by $Q = \pi r^2 (1-\cos\theta^*)\cos\theta^* c$. This counterintuitive hydraulic behavior results from the peristaltic shape changes of metaboly, and fundamentally differs from that of non-adherent polarized animal cells, which move like piston-like fixed-shape \emph{squirmers}.\cite{Bergert2015,Lighthill-1952-a} We then reasoned that, if crawling \emph{Euglena} cells in this limit were pumping fluid backwards, then hydraulic resistance could in fact act as a propulsive force.

To test theoretically this idea, we placed ourselves in the opposite limit of vanishing wall friction relative to hydraulic resistance, $\xi\rightarrow 0$, and computed cell velocity  by 
requiring zero induced flow rate as the bulge moves backwards. We found that in this scenario the pellicle slides relative to the wall. Furthermore, instead of stalling under high hydraulic resistance like polarized animal cells,\cite{Bergert2015}  peristaltic \emph{Euglena} cells actually move at a faster speed $v = -\left(1-\cos^2\theta^*\right)\, c$, Fig.~3(c)iii and Supplementary Movie~S9. In an intermediate regime, in which propulsion is governed by the balance of finite wall friction and hydrodynamic resistance, our model predicted that cells are propelled by hydraulic resistance and dragged by wall friction, opposite to polarized animal cells, Supplementary Note~4. In summary, despite seeming similarities between the crawling modes of \emph{Euglena} cells and of non-adherent polarized animal cells, our model highlighted fundamental differences associated with the large shape modulations of the former, and portrayed metaboly as a highly robust and adaptable mode of locomotion capable of using both frictional and hydraulic forces for propulsion depending on the mechanical nature of confinement. Furthermore, the fastest crawling cells in our experiments moved about 20 times faster than the fastest reported non-adherent polarized animal cells.\cite{Liu2015}

One of the most striking predictions of this model is that confined \emph{Euglena} cells, which act as hydraulic plugs, can move in a capillary without inducing any flow rate thanks to the shape deformations of metaboly. Supporting this prediction, we observed crawling cells accomplishing significant displacements next to stuck immobile cells, Fig.~3(d) and Supplementary Movie~S10. In the absence of stuck cells, we visualized the flow generated by crawling cells tracking the motion of suspended micron-sized beads, Fig.~3(e) and Supplementary Movie~S11. We found that beads in the vicinity of cells underwent rapid motions revealing  local flows induced by shape changes and flagellar beating. These bead velocities, however, exhibited rapid decay away from the cell body, where they were consistent with Brownian motion in a quiescent fluid.
Thus, in the light of our model, cells in our capillary experiments were close to the limit of high hydraulic resistance, which acted as the propulsive force. In agreement with this notion, we observed significant pellicle sliding in cells crawling and turning in a capillary, Fig.~3(f) and Supplementary Movies~S5 and S12. By contrast and showing the robustness of metaboly, our model suggests that cells between glass plates, where hydraulic confinement is very low, crawled by exploiting frictional propulsion, Fig.~1(b) and Supplementary Movies~S2 and S3.

\section*{Computational model}

Our previous observations have established that \emph{Euglena} cells develop highly concerted shape deformations to crawl under confinement, and that cells can adapt their gait to varying degrees of confinement between plates or in a capillary. We then tried to understand whether this complex adaptive behavior necessarily required a dedicated mechanosensing and mechanotransduction machinery or if, instead, the active pellicle could mechanically self-adapt its dynamics to a changing environment. To test this second possibility, we developed a theoretical and computational model of crawling in confinement by metaboly, which included the biological activity leading to strip sliding, the pellicle mechanics, and the interaction with the environment, Supplementary Note 5.

Despite similarities, the actuation mechanism and molecular players involved in metaboly are far less understood than those behind the much faster flagellar beating.\cite{Lineaar1968} On the basis of calcium precipitation assays and reactivation of metaboly in detergent-extracted cell models,  inter-strip sliding is thought to depend on molecular motors distinct from flagellar dyneins and associated to microtubules positioned along pellicular strips, which are locally and dynamically activated by the release/sequestration of cytoplasmic calcium from narrow subpellicular channels of endoplasmic reticulum.\cite{suzaki3,pellicle-volume} In our model, we assumed that the space-time activation pattern driving cell deformations was unaffected by varying confinement, consistent with the confinement-independent period of metaboly, Fig.~2(d). We modeled the pellicle as an extended motor system, and its activation as a space-time modulation of the sliding velocity between adjacent strips in the absence of force, $v_\textrm{motor}^0(s,t)$ where $s$ is arc-length along the strips and $t$ is time, Fig.~4(a)i,ii. Forces between adjacent strips in the sliding direction can affect the velocity of the motor system. We considered an affine relation between the force experienced by the motor system, $\tau$, and the actual sliding velocity, $v_\textrm{motor}$, Fig.~4(a)iii.\cite{Svoboda:1994aa,Abraham:2018aa} To determine the a priori unknown force distribution $\tau(s,t)$ acting on the extended motor system, we accounted for the pellicle elasticity and for its mechanical interaction with the environment, i.e.~the cellular pressure to maintain cell volume constant, hydraulic forces required to push the water column, and contact/frictional forces against the capillary, Fig.~4(a)iv. We designed the activity pattern, Fig.~4(a)ii, to match the typical shape dynamics of metaboly at low confinement.\cite{Arroyo2012} To solve the highly nonlinear equations governing the active mechanics of the deformable pellicle under confinement, we developed a spline-based finite element computational method.

In the absence of confinement, this model rapidly reached a limit cycle (or gait) within a few periods, exhibiting the characteristic peristaltic cell deformations of metaboly, Supplementary Movie~S13. As confinement incrementally constrained the shape excursions, we observed that the model self-adapted by developing a new limit cycle consistent with the imposed confinement, Fig.~4(b), which led to cell migration in the presence of frictional wall coupling and/or hydraulic resistance, see Fig.~4(c,d) for the limits of high hydraulic resistance and  high friction. Remarkably, the computational model reproduced all the features of confined crawling by metaboly reported in Fig.~2, including the kinematics of the gait under confinement, Fig.~4(b,c), and the non-monotonic relation between capillary diameter and cell velocity, Fig.~4(d), which we attributed to a tradeoff between the ability to develop large shape excursions at low confinement and the sustained contact of the bulge running along a longer cell at high confinement. In agreement with the theoretical model in Fig.~3(c), the computational model also predicted faster motion in the limit of high fluid resistance, in which the maximum velocity agreed quantitatively with our experiments. Taken together, these results support a view of the pellicle of euglenids as an elastic and extended motor system that, once biologically activated, can mechanically self-adjust to the degree of confinement to produce an effective gait.%

\section*{Outlook}

We have thus established that, in analogy with the mesenchymal-to-amoeboid transition of animal cells \cite{Liu2015} or the amoeboid-to-flagellar transition of \emph{Naegleria gruberi},\cite{Fritz-Laylin:2010aa} \emph{Euglena gracilis} develop a transition between flagellar and metaboly modes of locomotion triggered by confinement. Biophysically, the peristaltic movement of \emph{Euglena} during metaboly provides a new and remarkably robust mechanism of fast cell crawling. It is unclear, however, whether \emph{Euglena} take advantage of this capability in natural conditions, where they predominantly swim in the water column. How and why \emph{Euglena} cells retained an active pellicle, a vestige of a phagotrophic ancestry, and developed the ability to operate this machine in a non-reciprocal manner compatible with effective crawling is intriguing, as further emphasized by our examination of confined metaboly in a primary osmotrophic and a phagotrophic species of euglenids (Supplementary Note~6). From an engineering point of view, the ability of the pellicle to mechanically self-adapt and maintain the locomotory function under different geometric and mechanical conditions represents a remarkable instance of mechanical or embodied intelligence,\cite{pfeifer2007, kim2013} a design principle in bio-inspired robotics by which part of the burden involved in controlling complex behaviors is outsourced to the mechanical compliance of the materials and mechanisms that build the device.


\subsection*{Acknowledgements}
GN and ADS acknowledge the support of the European Research Council (AdG-340685-MicroMotility). MA acknowledges the support of the European Research Council (CoG-681434), the Generalitat de Catalunya (2017-SGR-1278 and ICREA Academia prize for excellence in research). We thank Stefano Guido for helpful discussions in the early stages of this study.

\subsection*{Author Contributions}
GN, MA and ADS conceived the study; AB provided cells and culture expertise; GN performed the experiments; GN, MA and ADS analyzed the data, performed theoretical analysis and wrote the paper.

\bibliographystyle{my_unsrtnat_bis}

\clearpage

\section*{Figures and captions}

\begin{figure}[h]
\centering
\makebox[\textwidth][c]{\includegraphics[width=1\textwidth]{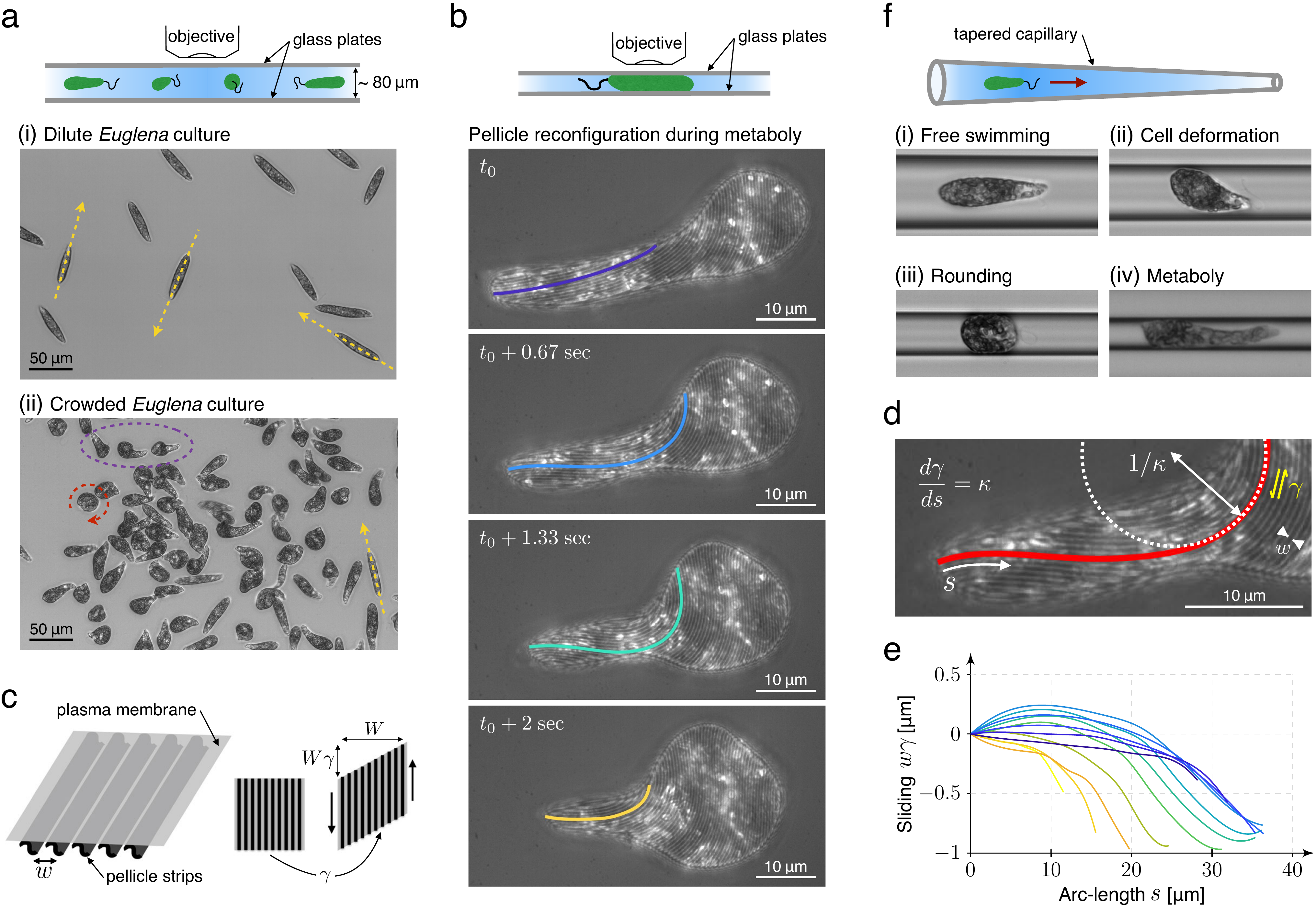}}
\caption{{\bf Confinement triggers changes in cell behavior and metaboly}. (a) \emph{Euglena} cells in dilute cultures (i) exhibited fast flagellar swimming without cell shape changes, at typical speeds of $68.2 \pm 1.13\,\upmu\text{m/sec}$ (SEM, n=50) or $1.31 \pm 0.03$ body lengths/sec (SEM, n=50). In crowded cultures (ii) cells displayed a variety of behaviors, including fast flagellar swimming (yellow arrow), cell rounding and spinning (red arrow), and large amplitude cyclic cell body deformations -- metaboly (purple ellipse), Supplementary Movie~S1. (b) When confined between glass plates, cells systematically developed metaboly. Observation using brightfield reflected light microscopy revealed the reconfigurations of the striated cell envelope (pellicle) concomitant with cell deformations in the plane of the glass plate (n=5 cells), Supplementary Movies~S2 and~S3. (c) Schematic representation of the pellicle, the strip separation $w$, and the local mechanism of active deformation of the pellicle, according to which the relative sliding between strips produces a shear strain $\gamma$ along the direction of the strips. During the shape dynamics in (b) $w$ remained nearly constant ($559 \pm 0.003\,\text{nm}$, SEM, n=100). (d) According to a mathematical model for the kinematics of the pellicle (Supplementary Note~1), the rate of change of $\gamma$ with arc-length $s$ along a strip (measured from the anterior end) is given by the curvature of the strip $\kappa$. By selecting strips in (b) emanating from the pole, where sliding is geometrically constrained, we could quantify the sliding displacement between adjacent strips $w\gamma$ required to bend an initially straight strip (e), where the color-coding matches the strips highlighted in (b). (f) Cells swimming into tapered capillaries transitioned from fast flagellar swimming (i) to developing large amplitude shape excursions (ii) including rounding (iii) and the prototypical highly orchestrated peristaltic cell deformations of metaboly (iv), Supplementary Movie~S4. This transition between (i) and (ii) occurred at a ratio between capillary and cell diameter of about $2.1\pm 0.05$ (SEM, n=10). }
\end{figure}
\clearpage

\begin{figure}[h]
\centering
\makebox[\textwidth][c]{\includegraphics[width=1\textwidth]{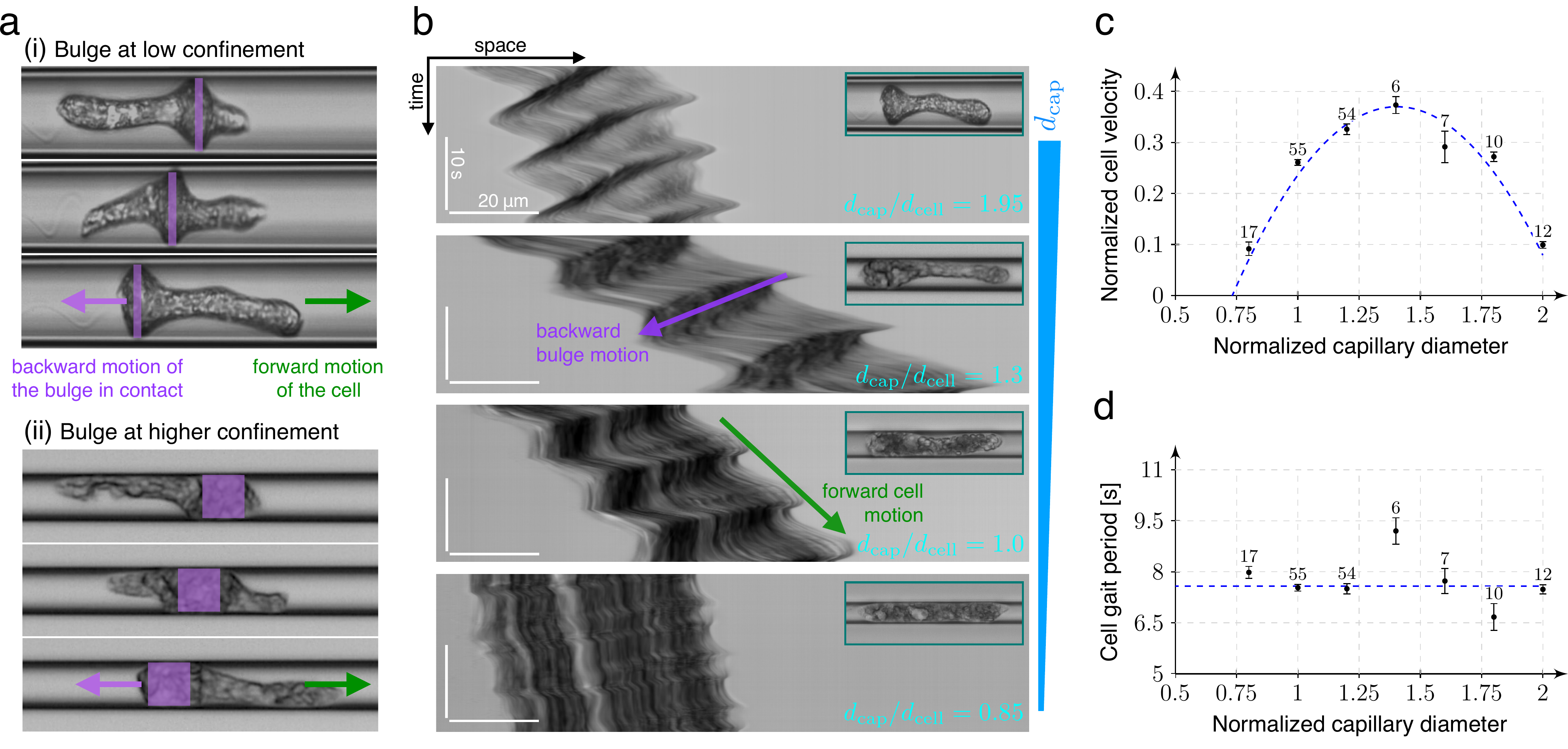}}
\caption{{\bf Metaboly is an effective crawling mode of locomotion under confinement}. (a) At low confinement, the bulge traveling backwards along the cell body transiently contacted the capillary walls, producing a net forward motion (i). At higher confinement, cells adapted the gait, with a larger contact area of the bulge and smaller amplitude of cell deformations. However, the crawling mechanism remained the same (ii). (b) Kymographs of crawling \emph{Euglena gracilis} under increasing confinement show the regularity of the gait, the backward bulge motion, and the forward cell motion. Crawling by metaboly was effective up to very large degrees of confinement (see the insets and Supplementary Movie~S6). The normalized velocity expressed in cell body lengths per period was maximum at an intermediated capillary diameter (c), whereas the period of the gait was largely insensitive to confinement (d). The blue dashed lines in (c) and (d) are guides to the eye. The capillary diameter $d_{\textrm{cap}}$ was normalized by the cell diameter $d_{\textrm{cell}}$ as it was free-swimming with a fixed cigar shape in the wider part of the capillary ($d_{\textrm{cell}} \simeq 9\,\upmu\text{m}$). The error bars in (c) and (d) refer to the standard error of the mean and the size of samples is indicated. Here one sample is a complete period, and data was gathered from 16 cells.}
\end{figure}
\clearpage

\begin{figure}[h]
\centering
\makebox[\textwidth][c]{\includegraphics[width=1\textwidth]{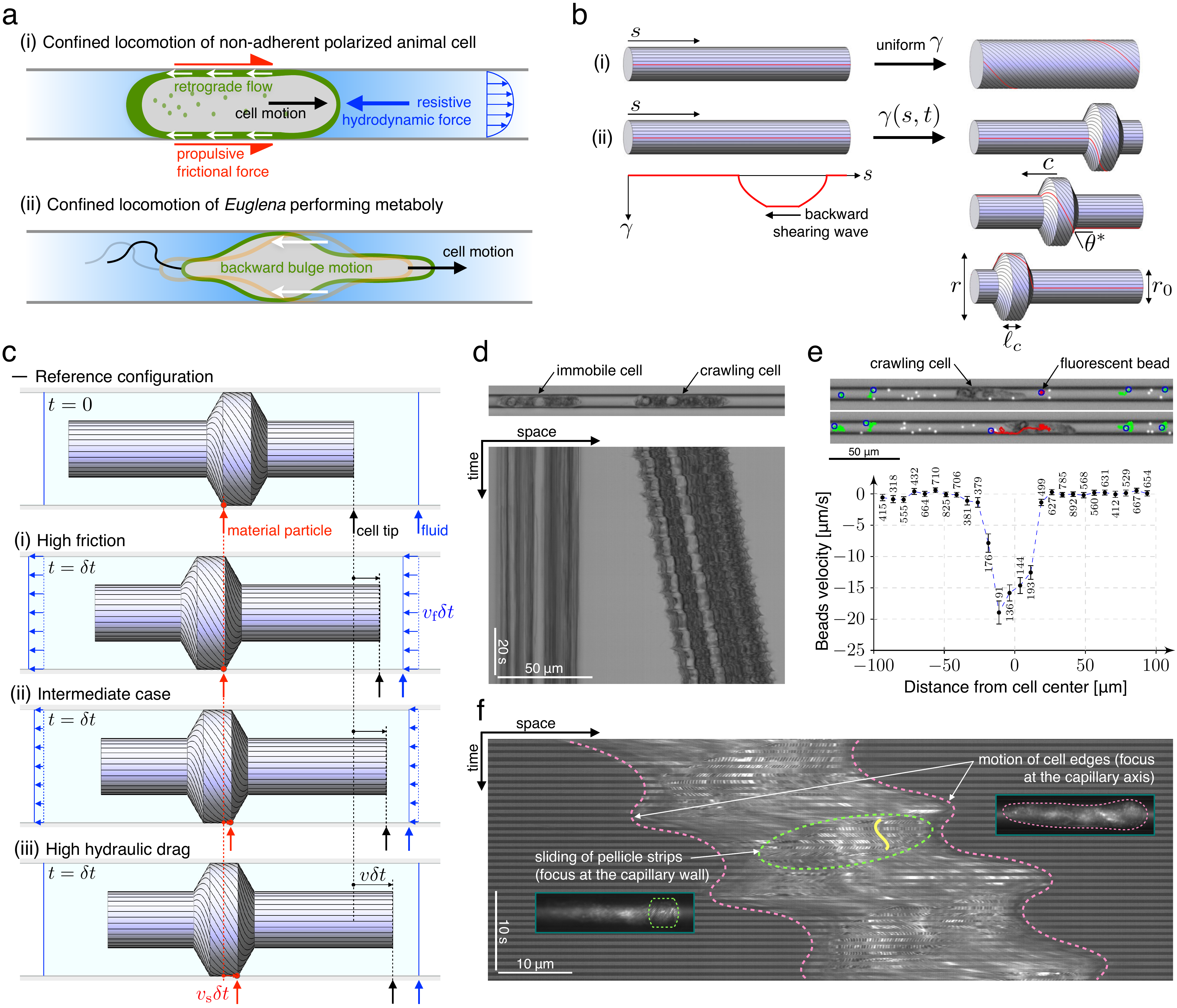}}
\caption{{\bf Mechanism of locomotion during metaboly}. (a) (i) Canonical model of propulsion of non-adherent polarized animal cells under confinement: frictional forces induced by actin retrograde flow propel the cell forward against resistive hydraulic forces required to displace the water column in the capillary. (ii) In \emph{Euglena} cells crawling by metaboly, the backward-moving pellicle bulge is analogous to actin retrograde flow in animal polarized cells. (b) Kinematics of the theoretical model for the power phase of metaboly. (i) Transformation of an idealized cylindrical pellicle by a uniform shear $\gamma$ along the strips. (ii) By propagating a pellicle shear profile $\bar{\gamma}(s)$ along the cell body following $\gamma(s,t) = \bar{\gamma}(s-c t)$, we model a moving localized bulge, which with our sign convention, moves leftwards at speed $c<0$ in the frame of the cell. (c) In the limit of infinite frictional coupling relative to hydraulic resistance (i), cell velocity is determined by the no-slip condition in the contact region. As indicated by the blue control volume, metaboly then results in net water pumping in the direction opposite to cell motion. The average fluid velocity $v_\textrm{f}$ is defined as the flow rate $Q$ divided by the cross-sectional area of the capillary. In the limit of zero frictional coupling relative to hydraulic resistance (iii), cell velocity is determined by the no water pumping condition, cells are fastest, and the pellicle slides in the contact region, as indicated by the red arrows. In intermediate cases (ii), hydraulic forces (propulsive) and frictional forces (resistive) compete, there is some degree of sliding and pumping, and the cell velocity is intermediate. (d) Kymograph of an immobile cell next to another cell crawling by metaboly, Supplementary Movie 10. (e) Quantification of fluid flow around crawling cells tracking suspended beads. Error bars refer to the standard error of the mean and sample size is indicated (a sampling point is the instantaneous velocity of a bead between two frames, data from 2 cells and 21 bead trajectories). (f) Kymograph made from images intermittently focused at the capillary wall to visualize the pellicle and at the capillary axis to visualize cell shape. The trajectories of pellicle features (yellow curve) reveal sliding between the pellicle and the capillary wall in the contact region.}
\end{figure}
\clearpage

\begin{figure}[h]
\centering
\makebox[\textwidth][c]{\includegraphics[width=1\textwidth]{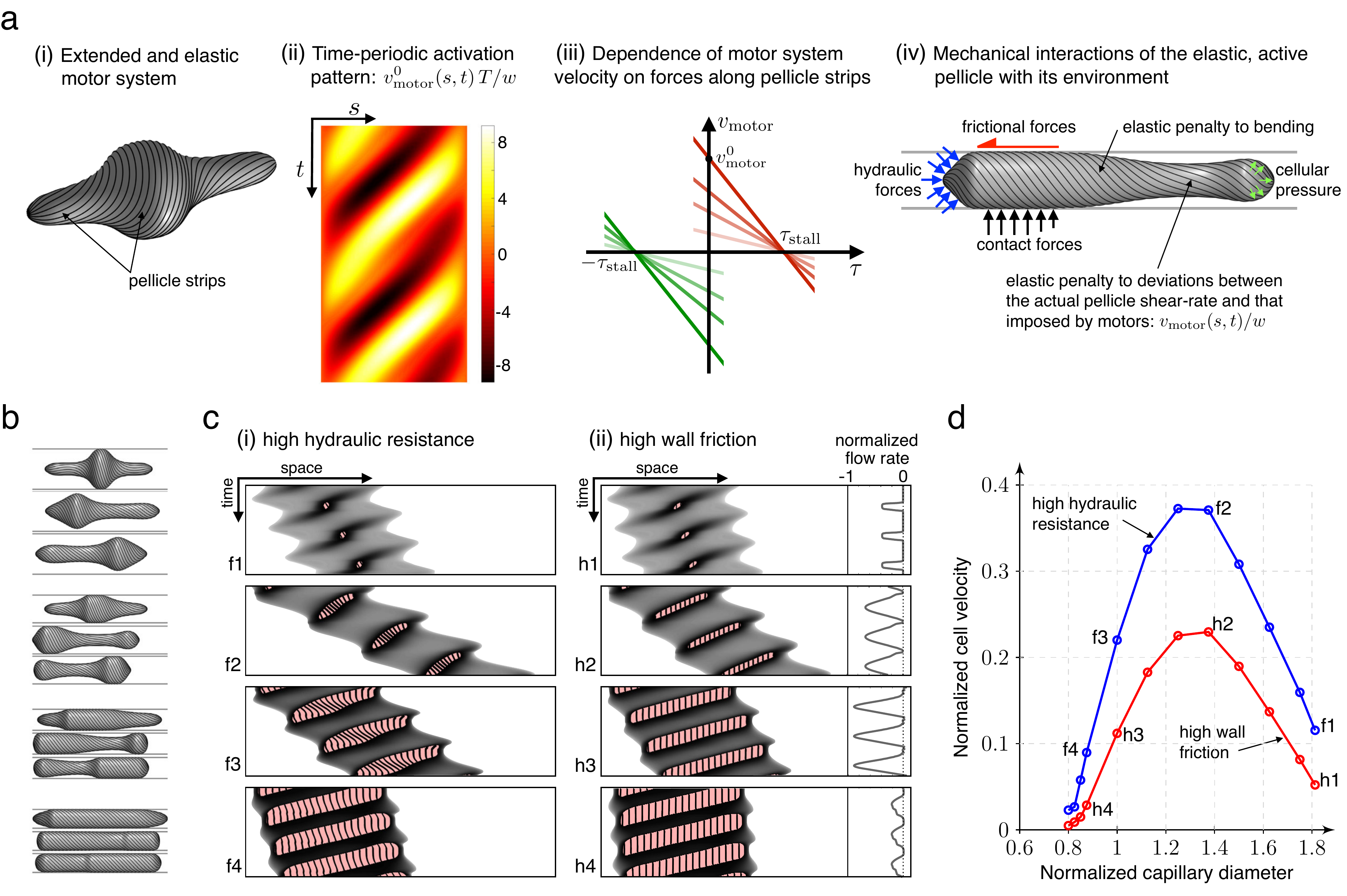}}
\caption{{\bf Computational modeling of crawling in confinement by metaboly}. (a) Model ingredients. (i) The pellicle is viewed as an elastic and extended motor system, modeled by an axisymmetric continuum surface with a field of material orientations converging towards two poles and accounting for the configuration of strips. The activation of this motor system in space (along the arc-length of strips, $s$) and time, $t$, is modulated to drive shape changes. Activation is modeled by the sliding velocity between adjacent strips in the absence of force, $v_\textrm{motor}^{0}(s,t)$. (ii) Time-periodic pattern of the activation $v_\textrm{motor}^{0}(s,t)$ during two gaits, in units of strip separation $w$ per period $T$. (iii) The actual sliding velocity $v_\textrm{motor}$ is modified by distributed forces along adjacent strips in the sliding direction, $\tau$. We model the force-dependent velocity of the motor system with an affine relation characterized by the time and space dependent $v_\textrm{motor}^0$ and by a stall force $\tau_\textrm{stall}$ of fixed magnitude, see Supplementary Note~5 for a discussion. (iv) The distributed force acting on the motor system, $\tau(s,t)$, is determined by the mechanical interactions of the pellicle with its environment (cellular pressure; contact, frictional and hydraulic forces) and by the pellicle mechanics, which include bending elasticity and stretching elastic forces that penalize deviations between the actual pellicle shear-rate and that imposed by the motion of motors, $v_\textrm{motor}(s,t)/w$. (b) Selected snapshots during the gait at four degrees of confinement. (c) Kymographs obtained from simulations in the high fluid resistance (i) and the high wall friction (ii) limits at the same four degrees of confinement. The contact region is colored in orange and the lines represent trajectories of material particles on the pellicle surface, showing sliding in the hydraulic-dominated case and no sliding in the friction-dominated case. Induced flow rate, normalized by the maximum cell velocity times the capillary cross-sectional area, is represented by gray curves in the case of high friction; it is zero in the limit of high hydraulic resistance. (d) Normalized cell velocity in body lengths per period as a function of normalized capillary diameter predicted by the simulations.
}
\end{figure}
\restoregeometry

\clearpage

\section*{Methods}
\subsection*{Culture of cells}

Strain SAG~1224-5/27 of \emph{Euglena gracilis} obtained from the SAG Culture Collection of Algae at the University of G\"ottingen (Germany) was maintained axenic in liquid culture medium Eg. Cultures were transferred weekly. Heterotrophic {\it Distigma proteus} and {\it Peranema trichophorum} were obtained from Sciento (https://www.sciento.co.uk). Subcultures of {\it P. trichophorum} were established in Eau Volvic with {\it Chlorogonium capillatum} (SAG~12-2b) as food source. All cells were kept in sterile 16\,mL polystyrene test tubes in an incubator IPP~110~plus from Memmert at 15\,\textdegree C and at a light:dark cycle of 12:12~h under cold white LED illumination with an irradiance of about 50\,$\upmu\text{mol}/(\text{m}^{2}\text{s})$.

\subsection*{Imaging of cells and preparation of tapered capillaries}

An Olympus BX\,61 upright microscope with motorized stage was employed in all experiments. These were performed at the Sensing and Moving Bioinspired Artifacts Laboratory of SISSA. Typically, the microscope was equipped with a LCAch~40X~Ph2 objective (NA\,0.55) for the imaging of cells behavior in capillaries using transmitted brightfield illumination. A Plan\,N~10X objective (NA=0.25) was employed to image cells and $1\,\upmu\text{m}$ in diameter polystyrene beads by combining brightfield and fluorescence microscopy (Supplementary Note~1 and Supplementary Movies~S7 and S11). The visualization of pellicle strips between microscope slides and in glass capillaries  was achieved by exploiting brightfield reflected light microscopy and using a UPlanFL\,N~100X objective (NA\,1.30~Oil). Micrographs were recorded with a CMOS digital camera from Basler (model acA2000-50gm) at a frame rate of either $20\,\text{fps}$ or $40\,\text{fps}$. The higher acquisition rate of $40\,\text{fps}$ was employed to time-resolve the dynamics of pellicle strips reconfigurations and for the study of wall friction as reported in Supplementary Note~2. Tracking of fluorescence beads was performed using Particle Tracker 2D/3D of the MosaicSuite for ImageJ.\cite{SBALZARINI2005182}
Tapered capillaries of circular cross section were obtained from borosilicate glass tubes (Sutter Instrument, model~B100-30-7.5HP) by employing a micropipette puller (Sutter Instrument, model P-97). A typical profile of such a capillary measured from micrographs is represented in Supplementary Fig.~2(b).
At each trial a glass capillary was filled with a diluted solution of cells in culture medium Eg and fixed to the microscope stage by means of a custom made, 3d-printed holder. To avoid optical aberrations that could arise from the curvature of the external surface of the capillary, this was positioned between two 0.17\,mm coverslips and covered with microscopy immersion oil.

\subsection*{Estimation of cell area and volume}

To estimate the area and volume of cells sandwiched between microscope glass slides, we measured the surface area $S$ and perimeter $P$ of the part of the cell surface in contact with the glass slides by focusing the microscope on these planes. We approximated the surface area of the cell envelope as $S_{\rm cell} = 2 S  + \pi H P/2$ and its volume by $V_{\rm cell} = SH  + \pi H^2 P /8 $, which assumes that cross-sections of the part of the cell surface not in contact with the plates are half circles. Using 15 frames from Supplementary Movie~S3, where $H = 4.3$ $\upmu$m, we estimated using this method $S_{\rm cell} \approx $ 1887\,$\upmu\text{m}^2$ $\pm$ 2\% and $V_{\rm cell} \approx $ 3283\,$\upmu\text{m}^3$ $\pm$ 1\%.

\subsection*{Code availability}

Mathematica (version 11.3.0.0) custom algorithms were developed and used to analyze the theoretical model in Fig.~3 and Supplementary Note~4. Matlab (R2017b) custom algorithms were developed and used to compute sliding displacements from strip curvature (Fig.~1(e) and Supplementary Note~1) and to implement the computational model in Fig.~4 and Supplementary Note~5. These computer codes are available from the corresponding authors upon reasonable request.

\subsection*{Data availability}

The data that support the plots within this paper and other findings of this study are available from the corresponding authors upon request.

\end{document}